\documentclass [12pt]{article}
\usepackage {epsfig}

\begin{document}

\begin{center}
{\Large {\bf Measurement of cosmic muon charge ratio with the Large Volume Detector}}

\vskip 0.5cm

{\bf 
N.Yu.~Agafonova$^{1,*}$, 
M.~Aglietta$^2$, 
P.~Antonioli$^3$, 
G.~Bari$^3$,  
R.~Bertoni$^2$, 
V.V.~Boyarkin$^1$,  
E.~Bressan$^{4,5}$,  
G.~Bruno$^6$, 
V.L.~Dadykin$^1$, 
E.A.~Dobrynina$^1$, 
R.I.~Enikeev$^1$, 
W.~Fulgione$^2$, 
P.~Galeotti$^{7,2}$, 
M.~Garbini$^3$, 
P.L.~Ghia$^8$,  
P.~Giusti$^3$, 
E.~Kemp$^9$, 
A.S.~Malgin$^1$, 
B.~Miguez$^{9,6}$,  
A.~Molinario$^2$, 
R.~Persiani$^{3,4}$, 
I.A.~Pless$^{10}$,  
V.G.~Ryasny$^1$, 
O.G.~Ryazhskaya$^1$,  
O.~Saavedra$^{7,2}$, 
G.~Sartorelli$^{3,4}$, 
M.~Selvi$^3$, 
G.C.~Trinchero$^2$,  
C.~Vigorito$^{7,2}$,  
V.F.~Yakushev$^1$, 
A.~Zichichi$^{3,4,5,11}$

(LVD Collaboration) 
} 
 
~~~

{\it
  
$^1$ Institute for Nuclear Research RAS, 117312,prospect 60-letya Oktyabrya,7a, Moscow, Russia

$^2$ INFN-Torino, OATO-Torino, 10100 Torino, Italy

$^3$ INFN-Bologna, 40126 Bologna, Italy

$^4$ University of Bologna, 40126 Bologna, Italy

$^5$ Centro Enrico Fermi, 00184 Roma, Italy

$^6$ INFN, Laboratori Nazionali del Gran Sasso, 67100 Assergi L.Aquila, Italy

$^7$ University of Torino, 10125 Torino, Italy

$^8$ Laboratoire de Physique Nucl.eaire et de Hautes Energies (LPNHE), Universit.es Paris 6 et Paris 7, CNRS-IN2P3, Paris, France

$^9$ University of Campinas, 13083-859 Campinas, SP, Brazil

$^{10}$ Massachusetts Institute of Technology, Cambridge, MA 02139-4307, USA

$^{11}$ CERN, Geneva, Switzerland

}
\end{center}

\noindent $^*$ Corresponding author: agafonova@lngs.infn.it 

\begin{abstract} 
 
The charge ratio ${k \equiv \mu^+/\mu^-}$ for atmospheric muons 
has been measured using Large Volume Detector  
(LVD) in the INFN Gran Sasso National Laboratory, Italy (minimal depth is 3000 m w.e.). To reach this  
depth muons should have the energy at the sea level greater than 1.3 TeV.  The muon charge ratio was defined 
using the number of the decays of stopping positive muons in the LVD iron structure and the  
decays of positive and negative muons in scintillator.  
We have obtained the value of the muon charge ratio ${k}$ ${= 1.26 \pm 0.04(stat) \pm 0.11(sys)}$. 
\end{abstract}


\vskip 0.5cm
 
\section{Introduction} 
\label{intro} 
The charge ratio of cosmic ray muons has been studied since the muon discovery. The muons  
charge ratio was used to obtain that of primary cosmic rays (p.c.r.) because the ratio 
${k \equiv \mu^+/\mu^-}$ at muon energy ${E_\mu\geq}$ 5 TeV depends on the p.c.r. charge ratio  
and on the characteristics of interactions of p.c.r. particles with air nuclei  
(differential cross-section of the ${\pi}$- and K-meson production, total cross-section).\\ 
The charge composition of near-vertical muon flux is of particular interest since  
near-vertical muons dominate in the total flux. The data obtained in 20  
experiments \cite{Ref_Nau},\cite{Ref_Bug} at the sea level up to ${P_\mu \sim}$100 GeV/c are 
in a good agreement with  
the standard concept about p.c.r spectrum and pA- and AA- interactions in the  
corresponding range of p.c.r. energy below 0.6 TeV.\\ 
There are very few measurements \cite{Ref_OPERA}, \cite{Ref_MINOS}, \cite{Ref_L3}, 
\cite{Ref_UTAH}, \cite{Ref_CMS} for the energy higher than 100 GeV.
The spread of the mean values of the ${\mu^+/\mu^-}$-ratio and significant errors of
the measurements do not allow to draw a conclusion about behavoir of the
 ${k}$ value at the energy $E_{\mu}{>}$1 TeV.\\ 
The new mechanisms of the secondary particle production in pA-interactions  
leading to the change of  the ${k}$ value can appear for $E_{\mu}{>}$ 100 GeV. 
The calculations  \cite{Ref_Vol} show that  
 the discovered quark-gluon state of matter (Quark Gluon Plasma - QGP) 
 in pA-interaction could lead to a progressive decrease of the $k$ value from  
${\sim}$1.3  at 100 GeV to  1.14 at 10 TeV. The $k$ value is of particular interest at 
the energies above 1 TeV where the number of available experimental data is quite low.\\
QGP is a state of matter at
large density of quarks and gluons. This high enery density state would be characterized by a strongly
reduced interaction between quarks and gluons so they would exist in a nearly free state. QGP was 
obtained at RHIC in BRAHMS, STAR, PHENIX and PHOBOS experiments 
in central collisions between gold nuclei at the center of mass energy of 100$A$GeV + 100$A$GeV.
The large number of articles produced by the four experiments at RHIC may be found on 
their respective homepages \cite{QGP}.
This energy is so large that conversion of a sizeable fraction of the initial kinetic energy into
matter production creates many thousands of particles in a limited volume.  \\
Nowadays, the LVD data can be used to obtain the positive excess of near-vertical  
muons of such energies and to evaluate the stopping muon charge composition \cite{Ref_LVD_pos}. The  
minimum depth of LVD location is 3000 m w.e., the average depth is 3600 m w.e. 
Only muons at energies above 1.3 TeV at surface 
can reach this depth. The muons, stopped at the LVD depth, have at surface the median energy 1.8 TeV. 
While 95\% of the stopping muons are within the range of 900 GeV - 3 TeV.
  The average flux of cosmic muons at the LVD depth
is ${(3.31 \pm 0.03) \times}$ ${10^{-4}(m^2s)^{-1}}$ \cite{Ref_LVD89}.
 \begin{figure}[!t] 
  \centering 
  \includegraphics[width=3.2in]{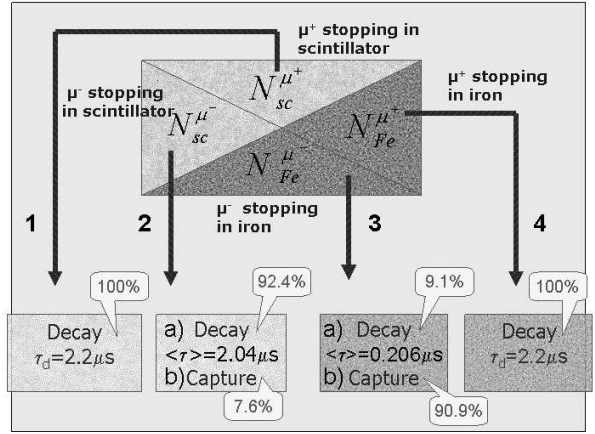} 
  \caption{The stopping muon processes in the LVD matter} 
  \label{1fig} 
 \end{figure} 
 
\section{Detection Method} 
 
LVD \cite{Ref_LVD} is located under Gran Sasso Mountain in central Italy. It is a  
scintillation-tracking detector with iron-hydrocarbonate target. The total detector mass is  
2 kt (iron and liquid scintillator). Scintillator ($C_nH_{2n}$, ${\overline n}$ = 9.6, 
with  ${\rho}$=0.78 ${g/cm^3}$    \cite{Ref_Dad}) and iron are  
distributed uniformly in the volume of the apparatus forming a modular structure  
of 840 elementary cells. The cells are grouped in three towers, each consisting  
of 7 layers. The tower dimension is 13 (length) ${\times}$ 6 (width) ${\times}$ 10 (height) $m^3$. 
The cell is a scintillation  
counter of a volume of $100 \times 100 \times 150$ $cm^3$ surrounded by 
iron, which has mean thickness  
of 2.9 cm. Each counter is viewed from the top by three photomultiplier tubes (PMTs) with diameter of 15 cm.
Eight counters are assembled into the iron module. The configuration of  
iron and scintillator permits to detect the products of nuclear interactions in iron  
using the scintillation counters. The detection energy threshold is 5 MeV.\\ 
The iron mass is equal to about 45\% of the total mass in the inner part of the tower.  
So the ratio of the muon stopping in the scintillator and in iron is \\ 
${N^{sc}_{st}/N^{Fe}_{st}=M_{sc}/M_{Fe} = 1.21}$ at the uncertainty of 2.5\%.

The ${\mu}$ stopping processes in the LVD materials are presented in Fig.~\ref{1fig}.\\  
 \begin{figure}[!t] 
  \centering 
  \includegraphics[width=3.5in]{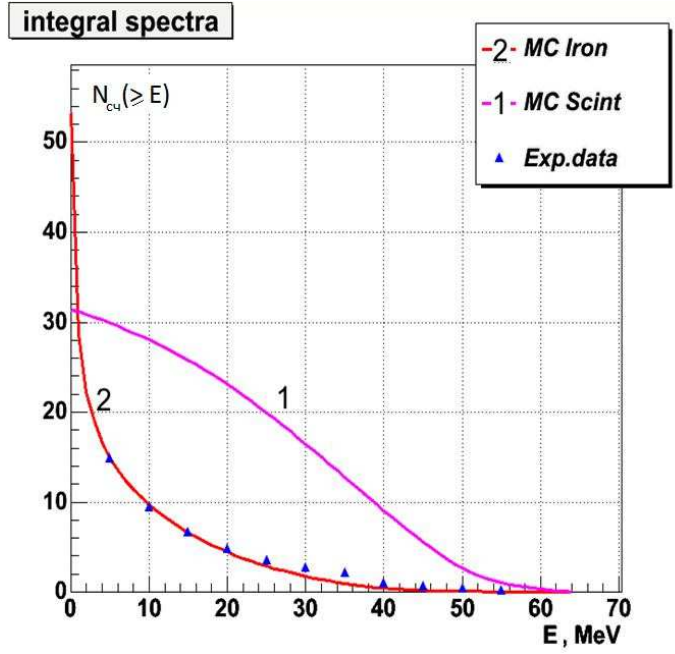} 
  \caption{The integral energy spectra of $\mu^\pm$  -decay events in scintillator (1) 
and ${\mu^+}$ ones in iron (2) per a counter. The curve (1) has
 been normalized by the total  
number of detected  ${\mu^\pm}$-decay events with energy ${>}$5 MeV. The curve (2) has been  
normalized by the experimental data of  ${\mu^+}$-decay in iron.} 
  \label{2fig} 
 \end{figure} 
The charge composition ${k}$ of the muon flux can be obtained by measuring the  
fraction of negative or positive muons in the total amount of stopping muons:  
\begin{equation} 
k=\frac{R^+}{R^-}=\frac{R^+_{sc}}{R^\pm_{sc}-R^+_{sc}} 
=\left(\frac{R^\pm_{sc}}{R^+_{sc}}-1\right)^{-1} 
\end{equation}   
where ${R^\pm_{sc}}$ and ${R^+_{sc}}$  are the numbers of ${\mu^\pm}$ and ${\mu^+}$  
stops in a counter normalized by the muon flux for the counter that is   
${R^\pm_{sc}=N^\pm_{sc}/N_\mu}$,${R^+_{sc}=N^+_{sc}/N_\mu}$. 
The features of LVD setup such as the configuration of iron and scintillator,  
the electronic dead time of 1 ${\mu s}$, the presence of photomultiplier afterpulses 
 allow to establish \\ 
{\it a)} the number of ${\mu^\pm}$ -decays in scintillator ${_dN^\pm_{sc}}$ ;\\ 
{\it b)} the number of ${\mu^+}$ -decays in iron  ${_dN^+_{Fe}}$ .\\ 
The value of {\it a)} is determined by searching for pulses in the  
scintillation counters crossed by muon, i.e. placed along the muon track. 
The value of {\it b)} can be obtained by using the data of the counters  
 that were adjacent to the counters crossed by muon. Using {\it a)} 
we can get the total number of  
stopping muons in scintillator. From {\it b)} we can go to ${_dN^+_{sc}}$ so, 
\begin{equation} 
_dN^+_{sc} \propto \frac{M_{sc}}{M_{Fe}} \cdot _dN^+_{Fe} 
\end{equation} 
 \begin{figure}
  \centering
  \includegraphics[width=3.5in]{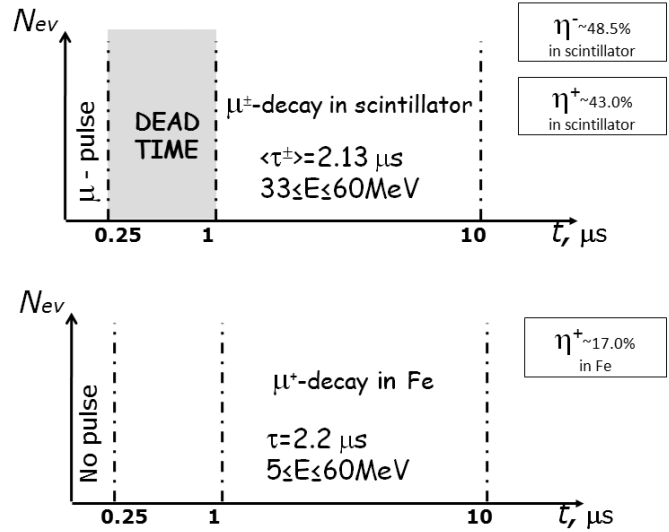}
  \caption{ 
The time diagrams for selection of the ${\mu}$-decay events. 
{\it The upper diagram:} selection of the ${\mu^\pm}$-decays in scintillator; the 
"${\mu}$-pulse" time interval contains pulses from counters crossed by 
muon, the "dead time" interval does not contain pulses due to the counter 
electronics dead time after muon pulse.
{\it The lower diagram:} selection of the ${\mu^+}$-decays in iron; the selected 
counters do not contain the muon pulses, the ${\mu^+}$-decay events are 
looked for in the time interval from 1 ${\mu}$s to 10 ${\mu}$s. 
The quantities ${\eta^- \sim}$ 48.5 \% and ${\eta^+ \sim}$ 43.0 \% are detection 
efficiencies for ${\mu^-, \mu}^+$ - decays in scintillator,  
${\eta^+ \sim}$ 17 \% is detection efficiency for ${\mu^+}$ -decays in 
iron. The detection efficiency values were obtained by MC simulation for 
the energy threshold of 5 MeV (Table 1).
}
  \label{3fig}
 \end{figure}
We can detect only ${\mu}$ decays but not captures. As follows from Fig.~\ref{1fig} number  
of ${\mu}$ decays in iron is less than number of ${\mu}$ stops.  
This fact is taken into account by using detection efficiency.\\

{\bf \em Stopping $\mu^+$} decays only.
\begin{equation} 
\mu^+ \rightarrow e^+ \nu_e \overline\nu_\mu, \tau_d = 2.2 {\mu}s \nonumber 
\end{equation}
The energy spectrum of ${e^+}$ has a maximum at ${\sim}$ 37 MeV and the greatest energy  of 52.8 MeV.\\ 
The positrons together with gamma-quanta from electromagnetic cascades  
(if muon decays in iron) and gamma-quanta from electron-positron annihilation  
are detected in the scintillation counters. The observed energy spectrum and 
 detection efficiency of ${\mu^+}$-decays in scintillator differ radically from  
the corresponding values of  ${\mu^+}$-decays in iron (Fig.~\ref{2fig}).\\

{\bf \em Stopping $\mu^-$} in detector materials may either decay or be captured by iron  
and carbon nuclei.
The energy spectrum of negative muon decay products is the same as for positive one,  
but the time characteristics depend on nuclear composition of the matter. \\
The rate ${\Lambda_c}$ of ${\mu^-A}$ -capture depends on ${Z}$ as ${Z^4}$ 
\cite{Ref_Primakoff}. Thus, in the case of  
stops in iron negative muons are mainly captured by iron nuclei (90.9\% of  
all  ${\mu^-}$-stops in iron), and in the case of stops in scintillator they  
mainly decay (92.4\% of all  ${\mu^-}$-stops in scintillator).  ${\mu^{- 12}C }$-captures 
 do not contribute to the final result because of the small fraction of 
 ${\mu^-}$-captures in the scintillator and of the small duration (1 ms) of the time 
 window for the data taking compared to the average lifetime of the products 
 of the reaction\\ 
 ${\mu^{-12}C \rightarrow ^{12}B\nu_\mu; ^{12}B \rightarrow ^{12}Ce^-\overline\nu_e }$ 
 (${\tau\sim39 ms}$). \\
The fraction of detectable muon captures in the  
scintillator is negligible (${\sim}$0.1\%). The probability of 
 ${\mu^-}$-capture by a free proton is 200 times less than the probability  
of  ${\mu^{- 12}C}$-capture, so this process was not taken into account.\\
${\mu^-Fe}$-capture is accompanied by gamma-quanta emission (0.32 gamma-quanta  
per capture) with energies of 3 - 10 MeV \cite{Ref_Vei} and also emission of 
${\sim}$1.13  
neutrons on average \cite{Ref_McD}. The time distribution of gamma-quantum pulses  
is described with a ${\mu^-}$ lifetime in iron: \\ 
${\tau_{Fe}=1/(\Lambda^-_c+\Lambda^+_d)}$ = 0.206
${\mu s}$ (at ${\Lambda^-_c = 44.0 \cdot 10^5 s^{-1}}$, 
${\Lambda^+_d = 4.52 \cdot 10^5 s^{-1}}$). $\Lambda^-_c$, 
$\Lambda^+_d$ are the rates of ${\mu}^-$-capture and ${\mu^+}$-decay in iron.\\
The same exponent corresponds to the  time distribution of ${\mu^-}$ -decays in iron.\\  
Although they represent the large fraction of events, ${\mu^-Fe}$-captures are not  
considered in our analysis because large amount of events in the time interval of 0.25 - 1.00 ${\mu s}$  
is the composition of the processes 3a, 3b,  
4 shown in Fig.~\ref{1fig} (See Fig.~\ref{3fig}).  
The range 0 - 0.25  ${\mu s}$ is used to define if a counter is crossed by muon or  
placed out the muon track. This range is a time interval of the counter hit delays due 
to delays of PMT responses.\\
  \begin{figure*}[th]
  \centering
  \includegraphics[width=5.9in]{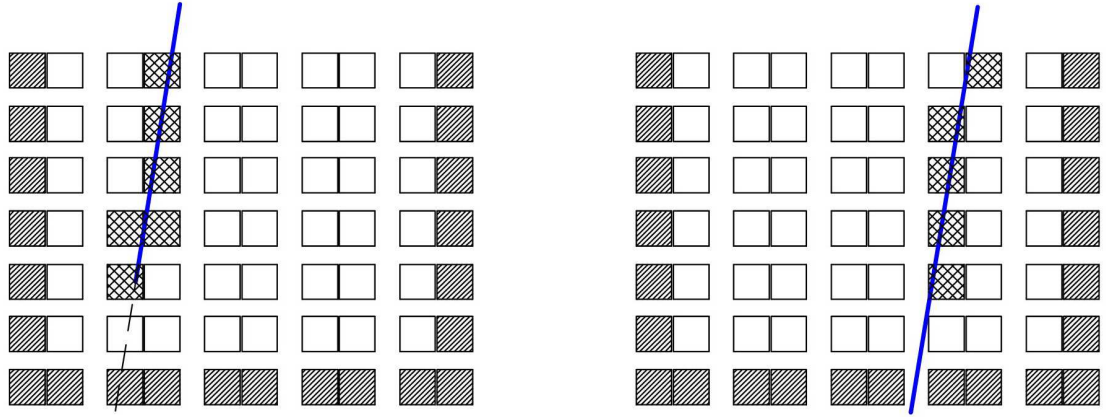}
  \caption{
The scheme of the LVD tower. Left - the stopping muon, 
a muon event is a cluster containing 6 muon pulses;
right - the quasi-stopping muon, a cluster contains 5 muon pulses.
}
  \label{LVD_f}
 \end{figure*}
\section{The selection criteria} 
We have analyzed the single muon events in the first LVD tower. A single muon event  
is defined as the presence of pulses with energy greater than 
5 MeV in several counters (from 2 to 11) in the time window 0.25 ${\mu s}$ from the  
first pulse of the cluster of pulses originated by muon (Fig.~\ref{LVD_f}). 
5 MeV is the threshold of detection of an energy released in a 
counter. The 
cluster corresponds to a single muon event. It is the 
set of muon pulses in counters crossed by muon. \\
Such a criterion  allows to eliminate the local 
produced muons, multiple muons and muons with accompanying shower. The local 
produced muons are arised from 
decays of pions generated in hadronic showers developed in the rock 
near detector. The energy of the local muons is not enough to cross more 
than one counter. Both multiple muons and muons with accompanying shower 
cross as a rule more than 11 counters.
The pulse at energy from 33 to 60 MeV in a  
time window of 1 - 10 ${\mu s}$ is regarded 
as candidate for ${\mu^\pm}$-decay   
in scintillator. The beginning of the time interval is determined by 
 dead time ${t_d}$ = 1.0 ${\mu s}$ after muon ionization loss pulse in a counter
(Fig.~\ref{3fig} upper diagram).\\ 
The counter time resolution for energy range corresponding to 
the energy deposit of a muon 
is ${\pm}$ 70 ns, TDC discreteness  
is 12.5 ns; the counter energy resolution is ${\sim}$25\% for 5 - 60 MeV energy 
range \cite{Ref_Amanda}.\\
The pulses in the same energy and time intervals but in the counters outside  
the muon track (i.e., without any pulse in the time window of 0.25 ${\mu s}$ )  
are considered as candidates for ${\mu^+}$-decays in iron. Due to 
 the ${\mu^-}$ lifetime in iron, the events of ${\mu^-Fe}$ -captures and ${\mu^-}$-decays 
have time less than 1 ${\mu s}$ (Fig.~\ref{3fig} lower diagram).\\
Detection efficiency for ${\mu^-Fe}$-capture in an 
adjacent counter is about $0.32 \cdot \eta_{\gamma}=0.32\cdot0.1=0.03$ (using 0.32 
gamma-quanta per capture). Here $\eta_\gamma$ is probability for gamma-quanta at energy of 1 - 3 MeV to imitate a 
pulse in an adjacent counter at energy higher of 5 MeV.  
Detection efficiency for ${\mu^-}$ -decay is 0.17 (see Fig.~\ref{3fig}, ${\eta^-_{Fe} =\eta^+_{Fe}}$). 
So, taking into account the ${\mu^-}$ decay fraction in Fe, which is about 9\%, 
we may neglect the number of negative muons that decay or get captured.\\
The data of 110 counters from 120 inner ones (from 2nd to 6th levels of first tower) 
 were used in the analysis. They were selected due to the time  
characteristics and stability. 
\section{Background estimation} 
The PMT afterpulses are the main background for the selection of ${\mu^\pm}$-decays  
in scintillator. The ${\gamma}$-quanta induced by ${nFe}$-captures are background  
 for ${\mu^+}$-decays in iron. Neutrons are also produced in ${\mu^-Fe}$-captures.\\
To exclude the background due to the PM afterpulses in the window of 1-10 ${\mu s}$, 
 all muon events are divided into 2 groups: through-going  
and ``quasi-stopping''. The muon is regarded as through-going  
if it produced a pulse in 0 - 0.25 ${\mu s}$   
time interval with energy higher than 10 MeV in the lower level counter.  
The remaining events are regarded as quasi-stopping (Fig.~\ref{LVD_f} right). Obviously,  
the ratio of muons really stopped  
in the LVD detector to the total amount of quasi-stopping muons is 
small because some muons can leave the  
detector through a gap (corridor) between columns (Fig.~\ref{LVD_f} left). Since 
there is no muon decay, the first group of events allows to determine the 
afterpulse  rate for each counter.\\
The number of real muon decays is thus given by the difference  between number of  
pulses in the 1 - 10 ${\mu s}$ window for the quasi-stopping events and the  
normalized number for through-going muons.\\
Then we get the integral time distribution in the 1 - 10 ${\mu s}$ window and total  
amount (at t = 0) of ${\mu^\pm}$ -decays in scintillator. At this stage we  
use ${\mu^\pm}$-decay exponent ${\tau = 2.2 \mu s}$ to eliminate  the background  
from ${nFe}$-captures which is almost flat in the range of 1 - 10 ${\mu s}$. \\
\section{Results} 
\label{sec:1} 
The analysis of the muon charge ratio is fulfilled using data of the  
1st LVD tower during 6 years of data collection. 
The data set contains 10986384 muon pulses in 110  counters. We have selected  
2299  ${\mu^\pm}$-decays in scintillator having an energy released by the  
decay products larger than 33 MeV. This additional cut was required  
to remove the events when muon crosses a counter, then stops and decays  
in iron wall while the decay products reenter into the same counter. 
 The energy release of the products of muon decay in iron does not exceed  
33 MeV which is approximately the mean value of the free muon decay spectrum (Fig~\ref{2fig}).\\ 
The number of  ${\mu^+}$-decays in iron is 1335. \\
\subsection{Normalized number of $\mu^\pm$ -decays in scintillator} 
 \begin{figure}[!h] 
  \centering 
  \includegraphics[width=3.5in]{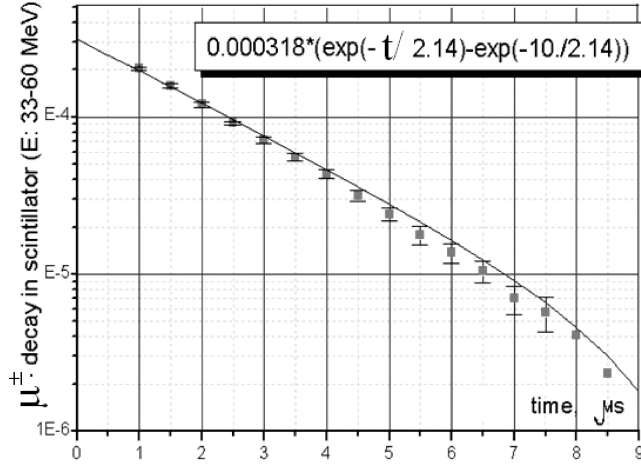} 
  \caption{The integral time distribution for $\mu^\pm$ -decays in scintillator. 
The curve is fit to the data at fixed exponent ${\tau}$ =2.14 ${\mu}$s. The bars represent the
statistical error.} 
  \label{5fig} 
 \end{figure} 
We obtain the value of ${R^\pm_{sc}}$  in (1) using the experimental  
number of  ${^{exp}R^\pm_{sc}=3.18 \cdot 10^{-4}}$  at the energy threshold of  
33 MeV and a corresponding detection efficiency ${\eta^\pm_{sc}}$: 
\begin{equation} 
R^\pm_{sc}=\frac{^{exp}R^\pm_{sc}}{\eta^\pm_{sc}} 
\end{equation} 
${^{exp}R^\pm_{sc}}$ is the value of the ${^{exp}R^\pm_{sc}(t)}$ function at  
t=0 (Fig~\ref{5fig}). Each point in the plot presents the  
value ${^{exp}R^\pm_{sc}}$  averaged 
over 110 counters at fixed ${t_{fix}}$ in the time interval of 1- 10 ${\mu s}$ 
\begin{displaymath} 
^{exp}R^\pm_{sc}(t_{fix})=\frac{\sum_{110}^{} (^{exp}R^\pm_{sc}(t_{fix}))}{110} 
\end{displaymath}
  \begin{figure}[!h] 
  \centering 
  \includegraphics[width=3.5in]{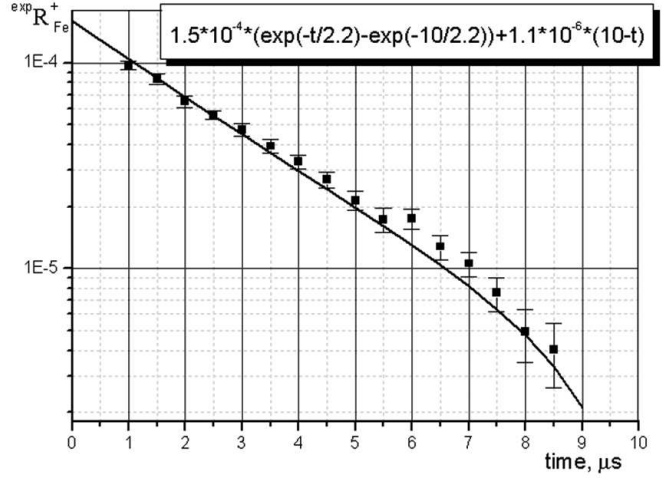} 
  \caption{The integral time distribution for $\mu^+$ -decays in iron. The
curve is the best fit to the data. The bars represent the statistical error.} 
  \label{6fig} 
 \end{figure} 
The value of  ${\eta^\pm_{sc}}$  depends on the detection efficiency  
(for energy threshold of 33 MeV) of ${\mu^\pm}$-decays, the charge  
composition of muons  ${c^+=\frac{k}{k+1}}$, ${c^-=\frac{1}{k+1}}$ and the  
fraction ${p^-_d}$  of ${\mu^-}$-decays in scintillator: 
\begin{equation} 
\eta^\pm_{sc}=c^+\eta^+_{sc}+c^-\eta^-_{sc}p^-_d, 
\end{equation} 
\setlength{\arraycolsep}{0.0em} 
\begin{eqnarray} 
p^-_d &{}={}& 1-p^-_d(^{12}C) \nonumber\\ 
      &{}={}& 1-\frac{\Lambda_c(^{12}C)}{\Lambda_c(^{12}C)+\Lambda_d(\mu^-)}=0.924  
\end{eqnarray}  
\setlength{\arraycolsep}{5pt} 
${\Lambda_c=4.52 \cdot 10^5}$, ${\Lambda_d=0.37 \cdot 10^5}$  are the rates 
of ${\mu^-}$-capture and ${\mu^-}$-decay  
in scintillator. Uncertainties in the rates are 2.7\% 
and 1.4\%,  respectively \cite{Ref_Vei}.
The values  of ${\eta^+_{sc}}$  and  ${\eta^-_{sc}}$  were calculated
 using a GEANT4 Monte Carlo simulation: ${\eta^+_{sc}}$ = 0.430,  ${\eta^-_{sc}}$ = 0.485 
(Table 1).
The efficiency ${\eta^+_{sc}}$ is less than ${\eta^-_{sc}}$ due to worser 
detection efficiency of gamma-quanta from ${e^+e^-}$-annihilation. Thus we get: 
\begin{equation} 
\eta^\pm_{sc}=\eta^+_{sc}\frac{k}{k+1}+p^-_d{\times}\eta^-_{sc}\frac{1}{k+1} 
\end{equation}  
To obtain the number of ${^{exp}R^\pm_{sc}}$ the experimental  
integral distribution (Fig.~\ref{5fig}) is fitted by the  
exponent having ${\tau^\pm}$ = 2.135 ${\mu s}$. Of course, the exponent depends  
on both the ${\mu^\pm}$ - lifetime in scintillator  
(${\tau^+}$ = 2.2 ${\mu s}$, ${\tau^-_{sc}}$  = 2.045 ${\mu s}$) and  
the muon flux charge composition but the dependence is weak: we have  
found that a variation of k from 1.0 to 1.5 changes the exponent  
 ${\tau^\pm}$  from 2.13 ${\mu s}$ to 2.14 ${\mu s}$. On the other  
hand,  ${^{exp}R^\pm_{sc}}$  varies insignificantly  
while changing ${\tau^\pm}$  value in the range 2.13 - 2.14 ${\mu s}$. 
 
\subsection{Normalized number of $\mu^+$ -decays in scintillator} 
We obtain the value of  $R^+_{sc}$  using the experimental value of   
$^{exp}R^+_{Fe} = 1.5 \cdot 10^{-4}$ (Fig.~\ref{6fig}) with provision for the mass  
factor  $M = M_{sc}/M_{Fe} = 1.21$ and the efficiency $\eta^+_{Fe}$.  Besides we need the quantity
${ B = \eta^\pm_b R^\pm_{sc}}$ which presents the amount of electrons ${e^\pm}$, penetrating into a given
 counter from an adjacent counter where ${\mu^\pm}$ decayed occured: 
\setlength{\arraycolsep}{0.0em} 
\begin{eqnarray} 
R^+_{sc} &{}={}& \frac{M}{a\eta^+_{Fe}}(^{exp}R^+_{Fe}-B) \nonumber\\
         &{}={}& \frac{M}{a \eta^+_{Fe}} (^{exp}R^+_{Fe}-\frac{\eta^\pm_b}{\eta^\pm_{sc}}\cdot ^{exp}R^+_{sc}) 
\end{eqnarray}  

\setlength{\arraycolsep}{5pt} 

The constant ${a}$ takes into account the fact that a counter detects ${\mu^+}$-decays  
in iron of an adjacent counter, thus ${a =2}$.
(Each side of an inner counter has only one side of an adjacent counter. 
The role of the contribution of the relative counter position in the ${\mu}^+$-decay detection 
 efficiency is taken into account by value of ${\eta}^+_{Fe}=0.17$).  
The ${\eta^\pm_b}$ coefficient is the fraction of ${\mu^\pm}$-decays in scintillator that  
produce pulses with energy larger than 5 MeV in an adjacent counter: 
$ 
\eta^\pm_b=c^+\eta^+_b+c^-\eta^-_bp^-_d. 
$ 
The values of ${\eta^+_b}$  and ${\eta^-_b}$  were calculated using GEANT4:  
 ${\eta^+_b}$ =0.06, ${\eta^-_b}$ =0.04 (Table 1).\\

\subsection{Determination of the charge ratio value} 

The ratio of ${R^\pm_{sc}/R^+_{sc}}$  in (1) is 
\begin{equation} 
\frac{R^\pm_{sc}}{R^+_{sc}}=\left(\frac{M\eta^\pm_{sc}}{2\eta^+_{Fe}}  
\cdot \left( \frac{^{exp}R^+_{Fe}}{^{exp}R^\pm_{sc}}-\frac{\eta^\pm_b}{\eta^\pm_{sc}}\right)\right)^{-1} 
\end{equation} 
and 
\begin{equation} 
k=\left\{\left( \frac{M \eta^\pm_{sc}}{2 \eta^+_{Fe}} \left( \frac{^{exp}R^+_{Fe}}{^{exp}R^\pm_{sc}} 
-\frac{ \eta^\pm_b}{\eta^\pm_{sc}}\right)\right)^{-1} -1 \right\}^{-1} 
\end{equation} 
The right part of the equation also depends on ${k}$. Thus we solve it numerically.\\
In conclusion, we have determined the single muon charge ratio for the flux of near-vertical muons in  
energy range from ${\sim}$ 1 TeV to ${\sim}$ 3 TeV at the sea level and  
average energy about 1.8 TeV: 
\begin{equation} 
k = 1.26 \pm 0.04(stat) \pm 0.11(sys). 
\end{equation}
\begin{figure}
\centering 
\includegraphics[width=4.0in]{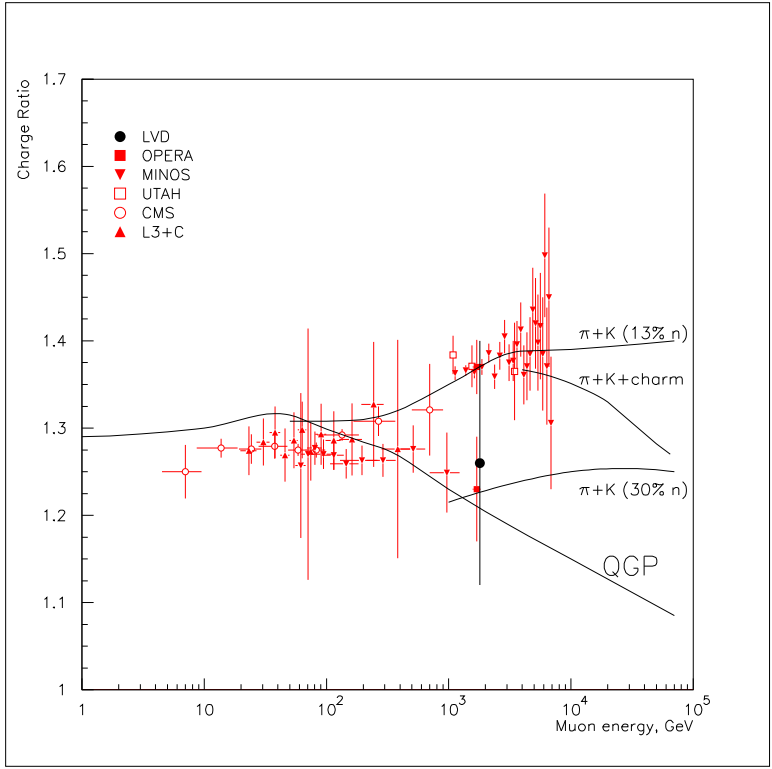}
\caption{The measured value of $k$ and the theoretical  
dependence $k(E_\mu)$ \cite{Ref_Vol}. The circle black point is the LVD result, the red points are
experimental data \cite{Ref_OPERA}, \cite{Ref_MINOS}, \cite{Ref_L3}, \cite{Ref_UTAH}, 
\cite{Ref_CMS}, the curves are  the theoretical predictions \cite{Ref_Vol}.} 
  \label{8fig} 
 \end{figure} 
\begin{table*}[th]
\centering
\caption{The systematic uncertainties of the quantities defining the muon charge ratio}
\label{table 1}

\begin{tabular*}{\textwidth}{@{\extracolsep{\fill}}llll@{}}
\hline
  &  Quantity & Magnitude  & Uncertainty, \% \\
\hline
           & $^{exp}R^+_{Fe}$   & 1.5 $\times 10^{-4}$  &  8  \\
 Measured  & $^{exp}R^\pm_{sc}$ & 3.18 $\times 10^{-4}$ &  5  \\
           & M                  & 1.21                  & 2.5 \\
           & ${p^-_d}$          & 0.924                 & 3 \\
\hline
                & $\eta^+_{Fe}$ & 0.17  & 2   \\
                & $\eta^+_{sc}$ & 0.430 & 1.5 \\
 Calculated (MC)& $\eta^-_{sc}$ & 0.485 & 1.5 \\
                & $\eta^+_b$    & 0.06  & 1.5   \\
                & $\eta^-_b$    & 0.04  & 1.5   \\
                & $\eta^{\pm}_b$   & 0.048 $(c^+ = c^- = 0.5)$ & 4 \\
\hline
\end{tabular*}
\end{table*}
The first error is statistic while the second one is systematic.
The statistic uncertainty depends on the statistic error of the quantities 
of ${^{exp}R^{\pm}_{Fe}}$ and ${^{exp}R^{\pm}_{sc}}$ in Eq.10. The values were obtained on basis 
of 1335 and 2299 events, correspondingly.  
The systematic error depends mainly on uncertainties of experimental values  
${^{exp}R^\pm_{sc}}$, ${^{exp}R^+_{Fe}}$, the mass factor ${M}$ and calculated numbers of  
${\eta_{Fe}}$, ${\eta_{sc}}$, ${\eta_b}$ (Table 1).\\
The systematic relative uncertainties in $^{exp}R_{sc}^{\pm}$ and $^{exp}R_{Fe}^+$ 
depend on the procedure of determinating of these 
values using the plotted points in Fig.~\ref{5fig},~\ref{6fig}. 
The error of the mass factor ${M}$ is 0.025. It is associated 
with accuracy of determination of scintillator and iron mass
which are, correspondlingly, 1150 $\pm$ 17 kg (${\sigma = 0.015}$) and 
950 $\pm$ 20 kg (${\sigma = 0.02}$) per a counter.\\
The uncertainty of the MC calculation of efficiency ${\eta^+_{Fe}}$
is about 2\% due to accuracy of determination of an iron module mass.
Uncertainty is the values of ${\eta^+_{sc}, \eta^-_{sc}, \eta^+_b, \eta^-_b}$
is 1.5 \% and it depends on accuracy of determination of the scintillator mass in a counter.
The uncertainty in the ${p^-_d}$ value depends on the measured rates of 
${\Lambda_c}$ (${\sigma}$ = 0.027) and ${\Lambda_d}$ (${\sigma =}$ 0.014) in Eq.6. It is 0.03.\\
Besides the known uncertainties in the values of ${\eta^+_b}$, ${\eta^-_b}$, ${p^-_d}$ the systematic error of
the ${\eta^{\pm}_b}$ value depends on uncertainty in the unknown charge composition ${c}$. As the ${\eta^{\pm}_b(c)}$ dependence 
is very week we can assume ${c^+ = c^- = 0.5}$ to estimate a value of ${\sigma(\eta^{\pm}_b)}$. At such a condition 
we have obtained ${\sigma(\eta^{\pm}_b) = 0.04}$.
The resultant systematic uncertainty is the square root of the quadrature sum of the errors of the values 
in the equation (10).\\
\section{Conclusion}

In LVD experiment the muon charge ratio was measured without the use of a magnetic field. Method of measurement 
is based on different probability of decay and capture of negative muons in heavy and light material. This method 
allows to measure the charge composition of the flux of single muons at the average muon energy of 270 GeV, which 
correspond to about 2 TeV at the surface.\\
Despite the large uncertainties, the LVD result supports  data in the energy range 0.2 - 2 TeV having  
the ${k}$ value of  about 1.28 (Fig.~\ref{8fig}). Perhaps this is due to the peculiarities of the charge 
composition of the near-vertical muon flux.\\

The experimental value of the muon charge ratio in comparisons 
with the theoretical predictions \cite{Ref_Vol} is shown in Fig.~\ref{8fig}. 
The first and third curves from the top of the figure are for muons from pion and kaon decays (neutrons
are 13\% and 30\% of protons in primary nucleon radiation). The second and forth curves are the results 
of calculations with the inclusion of charm (${\pi + K + charm}$) or QGP. \\
Given a set of the modern experimental data at the surface energies greater than 2 TeV 
obtained with magnetic spectrometer in the flux of non-vertical muons at zenith angles more than 45 degrees, one
 can conclude that, firstly, the QGP state does not 
play a significant role at pion production in ${pA, AA-}$ collisions 
in cosmic rays, and, secondly, the model included 30\% of neutrons is unrealistic.
\\
~~~

{\bf Acknowledgements.} 
 
This work is made with the RFFI financial support (grants 12-02-00213a, 
SSchool-3110.2014.2) and Program for Fundamental Research of Presidium of RAS.

\end{document}